\documentclass[preprint,prc,showpacs,preprintnumbers,amsmath,amssymb,floatfix]{revtex4}
\usepackage{amsmath}
\usepackage{lipsum}
\usepackage{graphicx}
\usepackage{dcolumn}
\usepackage{bm}
\usepackage{ulem} 
\usepackage[usenames]{color}
\usepackage{epstopdf}
\usepackage{epsfig}
\usepackage{float}
\usepackage{subfigure}
\usepackage{cancel}
\usepackage{multirow}
\usepackage[version=3]{mhchem}

\usepackage{soul}

\newcommand{\nc}{\newcommand}       
\nc{\vc}[1] {\mbox{\boldmath $#1$}} 
\nc{\del}       {\partial}              
\nc{\bra}       {\langle}               
\nc{\ket}       {\rangle}               
\nc{\bras}[1]   {\langle #1|}           
\nc{\kets}[1]   {|#1\rangle}            
\nc{\mapleft}[1]{           
	\smash{\mathop{\,          %
			\hbox to 1.5cm{\rightarrowfill}\, }\limits_{#1}}}
\nc{\beq}     {\begin{eqnarray}} \nc{\eeq}    {\end{eqnarray}}
\nc{\nn}      {\\\nonumber} \nc{\vs}      {\vspace{-0.275cm}}
\nc{\fra}    {\frac{1}{2}}
\nc{\mb}        {\mathbf}

\begin{document}
	
 \preprint{}
 
 \title{Density-dependent quark mean-field model for nuclear matter and neutron stars}
 \author{Kaixuan Huang}
 \affiliation{School of Physics, Nankai University, Tianjin 300071,  China}
 
 \author{Jinniu Hu}~\email{hujinniu@nankai.edu.cn}
 \affiliation{School of Physics, Nankai University, Tianjin 300071,  China}
 \affiliation{Shenzhen Research Institute of Nankai University, Shenzhen 518083, China}
 
 \author{Ying Zhang}
 \affiliation{Department of Physics, School of Science, Tianjin University, Tianjin 300354, China}

 \author{Hong Shen}~\email{songtc@nankai.edu.cn}
 \affiliation{School of Physics, Nankai University, Tianjin 300071,  China}

 \date{\today}
 
\begin{abstract}
      We develop a density-dependent quark mean-field (DDQMF) model to study the properties of nuclear matter and neutron stars, where the coupling strength between $\sigma$ meson and nucleon is generated by the degree of freedom of quarks, while other meson coupling constants are regarded as density-dependent ones. Two values for the nucleon effective mass, $M^*_{N0}/M_N=0.556,~0.70$ at the saturation density are chosen based on the consideration of the core-collapse supernova simulation and finite nuclei when the meson-nucleon coupling constants are fixed. We find that the equation of state (EOS) of nuclear matter, the symmetry energy, the mass-radius relations, and the tidal deformabilities of neutron stars with larger nucleon effective mass are more sensitive to the skewness coefficient $J_0$. The EOSs with  $M^*_{N0}/M_N=0.70$ are softer when the skewness coefficient $J_0=-800$ MeV. However, the maximum masses of the neutron star can be around $2.32M_\odot$ with $J_0=400$ MeV regardless of the value of the nucleon effective mass. By manipulating the coupling strength of the isovector meson to generate different slopes of symmetry energy, we construct the neutron star EOSs that can satisfy the different variables from the simultaneous mass-radius measurements of PSR J0030+0451, PSR J0740+6620 by the NICER collaboration, the mass-radius relations of  HESS J1731-347, and the radius constraints from the gravitational-wave signal GW170817 in the framework of DDQMF model. At the same time, most of these constructed EOSs can also satisfy the constraints of the tidal deformability from GW170817 event.
\end{abstract} 

 

\maketitle

\section{Introduction}\label{sec1} 
The properties of nuclear matter are strongly correlated with the matter in the core of heavy nuclei, neutron stars, and core-collapse supernovae, which are excellent testing grounds for studying nuclear many-body systems and have attracted many studies using different theoretical approaches, including the Brueckner-Hartree-Fock (BHF) approach \cite{jaminon1989}, the relativistic Dirac-Brueckner-Hartree-Fock (DBHF) approach \cite{brockmann1990}, the variational approach \cite{akmal1998} based on realistic nuclear forces, the chiral effective field theory (EFT) methods, as well as the Skyrme-Hartree-Fock, Gogny-Hartree-Fock \cite{bender2003}, and relativistic mean-field (RMF) models \cite{serot1986} based on the effective nucleon-nucleon ($NN$) interactions. 

All of the above models assume that the nucleons in the nuclear medium can be treated as point particles in the same way as those in free space. However, the EMC (European Muon Collaboration) effect indicates that the properties of the in-medium nucleon will be changed by the nuclear medium due to its internal structure \cite{aubert1983}, consisting of quarks and gluons. Nowadays, many experiments at different laboratories have been taken to investigate the structure of the nucleon. It is particularly worth mentioning that an electron-ion collider (EIC) is being built at Brookhaven National Laboratory to study the nucleon structure in finite nuclei precisely. Furthermore, the EIC in China (EicC) has been proposed and will be constructed based on an upgraded heavy-ion accelerator, High Intensity heavy-ion Accelerator Facility (HIAF) \cite{eicc2021}, which can offer significant insights into the three-dimensional landscape of the internal structure of the proton and other hadrons.

In addition, many theoretical works are devoted to studying the nuclear many-body theory from the quark level. Guichon proposed the quark-meson coupling (QMC) model \cite{guichon1988}, where the current quarks are confined in the MIT bag and the nucleons interact with each other through exchanging $\sigma$ and $\omega$ mesons between the quarks in different nucleons. Later, Toki et al. replaced the current quarks with constituent quarks in their proposed quark mean-field (QMF) model \cite{toki1998}, where the constituent quarks are confined by a confinement potential, and they applied the QMF model to study the properties of finite nuclei and neutron stars
\cite{shen2000,shen2002,hu2014prc,hu2014ptep}. Furthermore, to satisfy the spirit of QCD theory, Barik et al. developed a modified QMC model where the center of mass correction, pionic correction, and gluonic correction were taken into account when calculating the nucleon mass with the quark model \cite{barik2013, mishira2015, mishira2016}. Similarly, Xing et al. included the contribution of pions and gluons at the quark level within the QMF model \cite{xing2016} and applied it to the investigations of neutron stars and hypernuclei systems \cite{xing2017,hu2017,zhu2018,zhu2019,zhu2023,li2020J}. 

In these previous works, effective interactions for meson-nucleon couplings were based on the relativistic mean-field (RMF) approximation when performing nuclear matter calculations, including the nonlinear terms both for $\sigma$ and $\omega$ mesons \cite{shen2000,xing2016}, which can reproduce the nucleon self-energy from the DBHF theory \cite{brockmann1990} and satisfactory properties of finite nuclei \cite{sugahara1994}. In some sense, the meson self-coupling terms can be incorporated into the meson-nucleon coupling constants as a form of density dependence. Brockmann and Toki developed the density-dependent relativistic mean-field (DDRMF) method, where the coupling constants are density-dependent so that the corresponding self-energies are consistent with the DBHF results of nuclear matter \cite{brockmann1992}. 

Furthermore, the properties of neutron stars are more strongly dependent on the EOSs under extreme conditions of density and isospin asymmetry. With the rapid progress of astronomical-observable techniques, many works have focused on the observation and measurement of neutron stars, which can provide constraints on the EOS of neutron star matter. It is worth mentioning that the LIGO/Virgo collaborations have, for the first time, detected the gravitational wave produced from a binary neutron star merger, GW17087 \cite{abbott2017}, which provided crucial information about binary masses and the tidal deformability \cite{abbott2018}. The simultaneous measurements of mass-radius observations of the massive pulsars, PSR J0030+0451 \cite{miller2019,riley2019} and PSR J0740+6620 \cite{miller2021,riley2021}, can further provide constraints on the EOS. In addition to observations of massive neutron stars, a light central compact object in the supernova remnant HESS J1731-347 has recently been reported with a mass and radius of $M= 0.77^{+0.20}_{-0.17}M_\odot$ and $R = 10.4^{+0.86}_{-0.78}$ km, respectively \cite{doroshenko2022}. In our previous works \cite{huang2020,huang2022,huang2023}, the DDRMF model has been proven to be a very powerful many-body framework, which can describe the above observables very well.  

Therefore, we try to further develop a density-dependent quark mean-field (DDQMF) model, which incorporates the nuclear medium effects at the quark level, so as to study the EOS of dense matter and the properties of neutron stars. Unlike the DDRMF model \cite{typel1999}, the $\sigma$ meson-nucleon coupling constant does not have to be taken into account in the DDQMF model since the effective nucleon mass in the QMF model is obtained from quark level, where the $\sigma$ meson-quark coupling constant is introduced. As a result, the DDQMF model has fewer parameters than those in the DDRMF model. 
 
In this work, we aim to study the EOS of dense matter and the properties of neutron stars with the DDQMF model, where the constituent quarks ($m_q=350$ MeV) are confined by a potential in a harmonic oscillator form similar to Refs. \cite{barik2013, xing2016}. The density-dependent couplings for $\omega$ and $\rho$ mesons of the DDQMF model will be re-determined by fitting the saturation properties of nuclear matter from DDME-X model \cite{taninah2020}, which can reproduce the ground state properties of finite nuclei very well. The symmetry energy $E_{\rm sym}$ and its density dependence play a crucial role in the EOS of neutron star matter because of its highly isospin-asymmetric nature. $E_{\rm sym}$ and its slope ($L$) can be extracted from measurements of the neutron skin thickness ($R_{\rm skin}$) of $^{208}$Pb by PREX-II \cite{adhikari2021,reed2021,piekarewicz2021} and $^{48}$Ca by CREX collaboration \cite{adhikari2022}. However, the two measurements are very different, bringing a  great challenge to understanding the nuclear many-body theory. The slope of symmetry energy ($L$) can be controlled by adjusting the coupling constants of the isovector meson by fixing $E_{\rm sym}$ at the density of 0.11 $\rm fm^{-3}$ \cite{zhang2013}. We also tried to construct the EOS for neutron stars that can satisfy the observational constraints mentioned above using the DDQMF model.

The paper is organized as follows. In Sec. \ref{sec2} we briefly introduce the theoretical framework of the DDQMF model. In Sec. \ref{sec3}, the density-dependent parameters of meson couplings will be determined. The properties of nuclear matter and neutron stars obtained in the DDQMF models will also be shown. Finally, we will give a conclusion in Sec. \ref{sec4}.

\section{The Density-Dependent Quark Mean-Field Model}\label{sec2}
Within the QMF model, three constituent quarks are confined in the hadron by a confinement potential and satisfy the Dirac equation. After solving the Dirac equations in the presence of the meson mean fields, the effective mass of nucleons can be obtained, which will be used to solve nuclear many-body systems. In the nuclear medium, the Dirac equation for the constituent quarks can be written as
\begin{equation}\label{eqn.dirac1}  
 \left[i\gamma^{\mu}\partial_{\mu}-\left(m_q-g_{\sigma}^q\sigma\right)-\gamma^0\left(\Gamma_\omega^q\omega+\Gamma_\rho ^q\rho\tau_3\right)-U(\bm{r})\right]\psi_q(\bm{r})=0,
\end{equation}
where $\psi_q(\bm{r})$ represents the quark field with constituent quark mass $m_q$. $\sigma,~\omega,~\rho$ are the exchanging meson fields between quarks in different nucleons to achieve nucleon-nucleon interactions. $g_{\sigma}^q,~\Gamma_{\omega}^q,~\Gamma_{\rho}^q$ are the quark-meson coupling constants and $\tau_3$ is the third component of the isospin matrix. Here, we adopt a phenomenological confinement potential with a mixing scalar-vector form \cite{barik2013}, 
\begin{equation}\label{eqn.potential}
	U(\bm{r})=\frac{1}{2}\left(1+\gamma^0\right)\left(a_qr^2+V_q\right),
\end{equation}
since the analytical confinement potential for quarks cannot be obtained from QCD theory directly due to the highly nonperturbative at low energy. Now, the Dirac equation \eqref{eqn.dirac1} can be simplified as
\begin{equation}\label{eqn.dirac2}  
	\left[-i\bm{\alpha\cdot \nabla}+\beta m_q^*+U(\bm{r})\right]\psi_q(\bm{r})=\varepsilon_q^*\psi_q(\bm{r}),
\end{equation}
where 
\begin{equation} 
	\varepsilon_q^*=\varepsilon_q-\Gamma_{\omega}^q\omega-\Gamma_{\rho}^q\rho\tau_3,\quad
	m_q^*=m_q-g_{\sigma}^q\sigma,
\end{equation}
are the effective single quark energy and effective quark mass. Eq. \eqref{eqn.dirac2} can be solved exactly and its ground-state solution of the energy satisfies
\begin{equation}
	\label{eqn.condition}
	\sqrt{\frac{\lambda_q}{a_q}}\left(\varepsilon'_q-m'_q\right) =3.
\end{equation}
where 
\begin{gather}
	\varepsilon'_q=\varepsilon_q^*-V_q/2,\quad
	m'_q=m_q^*+V_q/2,\quad \lambda_q=\varepsilon'_q+m'_q=\varepsilon_q^*+m_q^*.
\end{gather}
The zeroth-order energy of the nucleon can be obtained from the solution of Eq. \eqref{eqn.dirac2} for the quark energy $\varepsilon_q$,
\begin{equation}
	E_N^*=\sum_q\varepsilon^*_q.
\end{equation} 
In this work, the center-of-mass correction $\epsilon_{\mathrm{c.m.} }$, the pion correction $\delta M_N^{\pi}$, and the gluon correction $\left(\Delta E_N\right)_{g}$ are taken into account following Refs. \cite{barik2013,xing2016}, so the mass of the nucleon in the nuclear medium becomes
\begin{equation}\label{eqn.nucmass}
	M_N^{*}=E_N^{* 0}-\epsilon_{\text {c.m.}}+\delta M_N^{\pi}+\left(\Delta E_N\right)_{g}.
\end{equation} 
 The specific form of each term in Eq. \eqref{eqn.nucmass} can be found in Ref. \cite{xing2016}. Finally, the nucleon radius in QMF model is written as
\begin{gather}\label{eqn.radius}
    \left\langle r_{N}^{2}\right\rangle 
    =\frac{11 \varepsilon_q^{\prime}+m_q^{\prime}}{\left(3 \varepsilon_q^{\prime}+m_q^{\prime}\right)\left(\varepsilon_q^{\prime 2}-m_q^{\prime 2}\right)} .
\end{gather}

Then we can apply the nucleon mass in the nuclear medium from Eq. \eqref{eqn.nucmass} to the nuclear many-body problem with the meson-exchange picture. To describe the nuclear matter, we consider the  scalar-isoscalar($\sigma$), vector-isoscalar ($\omega$), and vector–isovector ($\rho$) mesons and the DDQMF Lagrangian in the uniform system with mean-field approximation is given as
\begin{equation}  
	\mathcal{L}_{\mathrm{QMF}}=\bar{\psi}_N\left[i \gamma_{\mu} \partial^{\mu}-M_N^{*}-\gamma^{0}\left(\Gamma_{\omega N}(\rho_B)\omega +\Gamma_{\rho N}(\rho_B)\rho\tau_{3}\right)\right]\psi_N \\
	-\frac{1}{2} m_{\sigma}^{2} \sigma^{2}+\frac{1}{2} m_{\omega}^{2} \omega^{2}
	+\frac{1}{2} m_{\rho}^{2} \rho^{2},  
\end{equation}
where $\psi_N$ is the nucleon field. The effective nucleon mass, $M_N^*$ is obtained from the quark model as a function of the quark mass correction, $\delta m_q=-g_\sigma^q\sigma$, which is related to $\sigma$ field, while $\omega$ and $\rho$ mean fields do not obviously cause any change of the nucleon properties.  
The density-dependent coupling constant for $\omega$ meson can be expressed as a fraction of the baryon density, $\rho_B$, and the coupling constant for $\rho$ is chosen to be in exponential form,
\begin{equation}\label{eqn.ddcoup}
	\begin{aligned}
		\Gamma_{\omega N}(\rho_B)&=\Gamma_\omega(\rho_{B0})f_i(x),~~\text{with}~~f_\omega(x)=a_\omega\frac{1+b_\omega(x+d_\omega)^2}{1+c_\omega(x+d_\omega)^2},\\
		\Gamma_{\rho N}(\rho_B)&=\Gamma_\rho(\rho_{B0}){\rm exp}[-a_\rho(x-1)],
	\end{aligned}
\end{equation}
where $x=\rho_B/\rho_{B0}$ and $\rho_{B0}$ is the saturation density of symmetric nuclear matter. We keep the constraint $f_\omega(1)=1$, which can lead to
\begin{gather}\label{eqn.aw}
	a_\omega=\frac{1+c_\omega(1+d_\omega)^2}{1+b_\omega(1+d_\omega)^2},
\end{gather}
while the constraints $f''_i(0)=0$ and $f''_\sigma(1)=f''_\omega(1)$ in conventional DDRMF model \cite{typel1999} do not need to be considered here.

The equations of motion of nucleons and mesons will be generated by the Euler-Lagrangian equation,
\begin{equation}
	\begin{gathered} 
		 \left[i\gamma^{\mu}\partial_{\mu}-M_N^*-\gamma^0\left(\Gamma_{\omega N}(\rho_B)\omega+\Gamma_{\rho N}(\rho_B)\rho\tau_{3}+\Sigma_R\right)\right]\psi_N=0, \\
		m_{\sigma}^2\sigma=-\frac{\partial M_N*}{\partial\sigma} \rho_s,\\
		m_{\omega}^2\omega = \Gamma_{\omega N}(\rho_B)\rho_B,\\
		m_{\rho}^2\rho=\Gamma_{\rho N}(\rho_B)\rho_{B3},
	\end{gathered}
\end{equation}
where the $\frac{\partial M_N*}{\partial\sigma}$ is not a explicit function of the $\sigma$ mean field while that is equal to the $\Gamma_{\sigma N}$ in the DDRMF model \cite{huang2020}. The rearrangement term, $\Sigma_R$, is 
\begin{equation}
	\Sigma_R= \frac{\partial\Gamma_{\omega N}(\rho_B)}{\partial\rho_B}\omega\rho_B
	+\frac{\partial\Gamma_{\rho N}(\rho_B)}{\partial\rho_B}\rho\rho_{B3},
\end{equation}
where the scalar, vector densities, and their isospin components are generated by the expectation value of nucleon fields,
\begin{equation}
   \rho_s = \langle\bar{\psi}\psi\rangle,\quad \rho_B=\langle\psi^\dagger\psi\rangle,\quad \rho_{B3}=\langle\bar{\psi}\tau_3\gamma^0\psi\rangle.
\end{equation}
 
With the energy-momentum tensor, the energy density, $\mathcal{E}$, and pressure, $P$, of nuclear matter can be obtained respectively as
\begin{equation}
  \begin{aligned}
	\mathcal{E}=&\frac{1}{2}m_{\sigma}^2\sigma^2-\frac{1}{2}m_{\omega}^2\omega^2-\frac{1}{2}m_{\rho}^2\rho^2+\Gamma_{\omega N}(\rho_B)\omega\rho_B+\Gamma_{\rho}(\rho_B)\rho\rho_{B3}+\mathcal{E}_{\rm kin}^n+\mathcal{E}_{\rm kin}^p,\\
	P=
	&\rho_B\Sigma_{R}(\rho_B)-\frac{1}{2}m_{\sigma}^2\sigma^2+\frac{1}{2}m_{\omega}^2\omega^2+\frac{1}{2}m_{\rho}^2\rho^2+P_{\rm kin}^n+P_{\rm kin}^p,
  \end{aligned}
\end{equation}
where the contributions from kinetic energy are 
\begin{align} 
	\mathcal{E}_{\rm kin}^i&=\frac{1}{\pi^2}\int_{0}^{k_{Fi}}k^2\sqrt{k^2+{M_N^{*}}^{2}}dk
	=\frac{1}{8\pi^2}\left[k_{Fi}E_{Fi}^{*}\left(2k_{Fi}^2+{M_N^{*}}^2\right)+{M_N^{*}}^4{\rm ln}\frac{M_N^{*}}{k_{Fi}+E_{Fi}^{*}}\right], \\\nonumber 
	P_{\rm kin}^i&=\frac{1}{3\pi^2}\int_{0}^{k_{Fi}}\frac{k^4 dk}{\sqrt{k^2+{M_N^{*}}^{2}}}
	=\frac{1}{24\pi^2}\left[k_{Fi}\left(2k_{Fi}^2-3{M_N^{*}}^2\right)E_{Fi}^{*}+3{M_N^{*}}^{4}{\rm ln}\frac{k_{Fi}+E_{Fi}^{*}}{M_N^{*}}\right].
\end{align} 

Then, the properties of nuclear matter can also be determined. The binding energy per nucleon, $E/A$,  the incompressibility, $K$, and the skewness coefficient, $J$,  are defined by \cite{dutra2014}
\begin{align} 
	\label{eqn.ea}
	\frac{E}{A}=&\frac{\mathcal{E}}{\rho_B}-M_N,\\
	\label{eqn.K}
	K=&9\frac{\partial P}{\partial \rho_B}\bigg|_{\delta=0}
	=9\left[\rho_B\frac{\partial \Sigma_R}{\partial \rho_B}+\frac{2\Gamma_{\omega N} \rho_B^2}{m_\omega^2} \frac{\partial \Gamma_{\omega N}}{\partial \rho_B}
	+\frac{\Gamma^2_{\omega N}\rho_B}{m_\omega^2}+\frac{k_F^2}{3E_F^*}+\frac{\rho_B M_N^*}{E_F^*} \frac{\partial M_N^*}{\partial \rho_B}\right],\\
	\label{eqn.J}
	J=&\left.27 \rho_B^3 \frac{\partial^3\left(\mathcal{E}  / \rho_B\right)}{\partial \rho^3}\right|_{\delta=0}
	=27 \rho_B^3\left[\frac{1}{\rho_B} \frac{\partial^3 \mathcal{E} }{\partial \rho_B^3}-\frac{3}{\rho_B^2} \frac{\partial^2 \mathcal{E}}{\partial \rho_B^2}+\frac{6}{\rho_B^3} \frac{\partial \mathcal{E} }{\partial \rho_B}-\frac{6 \mathcal{E} }{\rho_B^4}\right],
\end{align} 
where
\begin{equation} 
\begin{aligned}
	\frac{\partial \mathcal{E} }{\partial \rho_B}= & \sqrt{k_F^2+M_N^{* 2}}+\frac{\Gamma_{\omega N}^2}{m_\omega^2} \rho_B+\Sigma_R, \\
	\frac{\partial^2 \mathcal{E} }{\partial \rho_B^2}= & \frac{1}{2 E_F^*}\left(\frac{\pi^2}{k_F}+2 M_N^* \frac{\partial M_N^*}{\partial \rho_B}\right)+\frac{\Gamma_{\omega N}^2}{m_\omega^2}+\frac{2 \Gamma_{\omega N} \rho_B}{m_\omega^2} \frac{\partial \Gamma_{\omega N}}{\partial \rho_B}+\frac{\partial\Sigma_R}{\partial \rho_B},  \\
	\frac{\partial^3 \mathcal{E} }{\partial \rho_B^3}= & -\frac{1}{4 E_F^{* 3}}\left(\frac{\pi^2}{k_F}+2 M_N^* \frac{\partial M_N^*}{\partial \rho_B}\right)^2+\frac{1}{2 E_F^*}\left[-\frac{\pi^4}{2 k_F^4}+2\left(\frac{\partial M_N^*}{\partial \rho_B}\right)^2+2 M_N^* \frac{\partial^2 M_N^*}{\partial \rho_B^2}\right] \\
	& +\frac{2 \Gamma_{\omega N}\rho_B}{m_\omega^2} \frac{\partial^2 \Gamma_{\omega N}}{\partial \rho_B^2}+\frac{2 \rho_B}{m_\omega^2}\left(\frac{\partial \Gamma_{\omega N}}{\partial \rho_B}\right)^2+\frac{4 \Gamma_{\omega N}}{m_\omega^2} \frac{\partial \Gamma_{\omega N}}{\partial \rho_B}+\frac{\partial^2 \Sigma_R}{\partial \rho_B^2}.\nonumber
\end{aligned}
\end{equation}
The symmetry energy, $E_{\rm sym}$, and its slope, $L$, are
\begin{align}
	\label{eqn.esym}
	E_{\rm sym}=&\frac{1}{2}\frac{\partial^2\varepsilon/\rho_B}{\partial\rho_B}\bigg|_{\delta=0}
	=\frac{k_F^2}{6E_F^*}+\frac{{\Gamma^2_{\rho N}(\rho_B)}}{8m_{\rho}^2}\rho_B,\\
	\label{eqn.L}
	L=&3\rho_{B}\frac{\partial E_{\rm sym}}{\partial\rho_B}=\frac{k_F^2}{3E_F^{*}}-\frac{k_F^4}{6E_F^{*3}}\left(1+\frac{2M_N^{*}k_F}{\pi^2}\frac{\partial M_N^{*}}{\partial\rho_B}\right)+\frac{3\Gamma_{\rho N}^2}{8m_{\rho}^2}\rho_B+
	\frac{-3a_\rho\Gamma_{\rho N}^2}{4m_{\rho}^2\rho_{B0}} \rho_B^2.
\end{align} 
We can find that $E_{\rm sym}$ and $L$ are only dependent on $\Gamma_{\rho N}(\rho_B)$ when the isoscalar properties of nuclear matter at nuclear saturation density are fixed. In addition, we define the scalar potential, $U_S$, and vector potential, $U_V$, as
\begin{equation}\label{eqn.UsUv} 
	U_S=M_N^*-M_N,\quad
	U_V=\Gamma_{\omega N}(\rho_B)\omega +\Gamma_{\rho N}(\rho_B)\rho\tau_{3}. 
\end{equation}
For comparison with the DDRMF model, we define the coupling constants of $\sigma$ meson, $\Gamma_{\sigma N}(\rho_B)$, with $M_N^*=M_N-\Gamma_{\sigma N}(\rho_B)\sigma$ in DDQMF model.

In the uniform neutron star matter, the compositions of baryons and leptons are determined by the requirements of charge neutrality and $\beta$-equilibrium conditions,
\begin{equation}\label{eqn.betaequip}
            \mu_\mu=\mu_e=\mu_n+\mu_p,\quad \rho_p=\rho_e+\rho_\mu.
\end{equation}

\section{Results and Discussions}\label{sec3}
The two parameters of the confinement potential, $(a_q,~V_q)$, are fixed to $(0.4955927,~ -102.041429)$ respectively for $m_q=350$ MeV by reproducing the experiment data of nucleon mass $M_N=939$ MeV and the charge radius $\left\langle r^2_N\right\rangle^{1/2}=0.87~\rm fm$ in free space with Eq. \eqref{eqn.condition}, Eq. \eqref{eqn.nucmass}, and Eq. \eqref{eqn.radius}. The effective mass $M_N^*$ is dependent on $\delta m_q=m_q-m_q^*=g^q_\sigma\sigma$ and it can be expanded in terms of $\sigma$ field to the fourth-order in symmetric nuclear matter,
\begin{equation}
	M_N^* = M_N+a(g_\sigma^q\sigma)+b(g_\sigma^q\sigma)^2+c(g_\sigma^q\sigma)^3+d(g_\sigma^q\sigma)^4.
\end{equation}
where the parameters $a=-2.19849,~ b=1.09324\times 10^{-3},~c=-6.20770\times 10^{-7},~d=8.47995\times 10^{-9}$ can be determined by fitting to the results of $M_N^*$ from Eq. \eqref{eqn.nucmass}.

The quark-meson coupling $g_{\sigma}^q$, the coupling parameters $\Gamma_{\omega N}(\rho_{B0}),~b_{\omega},~c_{\omega},~d_{\omega}$ and $\Gamma_{\rho N}(\rho_{B0}),~a_{\rho}$ can be determined by fitting saturation properties of nuclear matter, i.e., the saturation density, $\rho_{B0}$, the binding energy per nucleon, $E/A$, the incompressibility, $K_0$, the skewness coefficient, $J_0$, the effective mass, $M_{N0}^*/M_N$, the symmetry energy, $E_{\rm sym 0}$, and its slope, $L_0$, at the saturation point. The saturation properties used in this work are listed in Table \ref{table.SNMsat}, which are almost extracted from DDME-X set \cite{taninah2020}. $J_0$ at the saturation point is only loosely known to be in the range of $-800\leq J_0\leq 400$ MeV based on the analysis of terrestrial nuclear experiments and energy density functional \cite{zhang2018}. Here we choose another nucleon effective mass of $M_{N0}^*/M_N=0.70$ at the saturation point since a larger effective mass can lead to a more rapid contraction of the proto-neutron star, which will directly result in a faster explosion in the core-collapse supernova simulations \cite{yasin2020}, while $M_{N0}^*/M_N\sim 0.60$ can give reasonable spin-orbit splittings for finite nuclei in RMF model. 

\begin{table}[H] 
	\centering 
	\caption{Saturation properties, i.e. the saturation density, $\rho_{B0}$, the binding energy per nucleon, $E/A$, the incompressibility, $K_0$, the skewness coefficient, $J_0$, the effective mass, $M_{N0}^*/M_N$, the symmetry energy, $E_{\rm sym 0}$ and the slope of the symmetry energy, $L_0$, at the saturation point, used in this work for fitting the meson coupling constants.}\label{table.SNMsat}
	\begin{tabular}{cccccccccc}
		\hline \hline
		$\rho_{B0}$ [$\rm fm^{-3}$] ~~&$E/A$ [MeV] ~~&$K_0$ [MeV] ~~&$J_0$ [MeV] ~~&$E_{\rm sym 0}$ [MeV]  ~~&$L_0$ [MeV] ~~&$M_{N0}^*/M_N$ \\
		\hline    
		0.152  & -16.1  & 267  &-800/400  & 32.3   & 49.7   & 0.556/0.70  \\
		\hline\hline
	\end{tabular}
\end{table}
With $M_{N0}^*$, $\rho_{B0}$ and the correspondence between $M_N^*$ and $\delta m_q$, $g_{\sigma}^{q}$ and $\sigma$ field at the saturation point, $\sigma_0$, can be calculated,
\begin{gather} 
	g_\sigma^q={-\frac{m_\sigma^2\delta m_q}{\frac{\partial M_{N0}^*}{\partial\delta m_q }\rho_s}},\\
	\sigma_0=\frac{\delta m_q}{g_\sigma^q}.
\end{gather}
Along with the $E/A$ at the saturation point, the parameters, $\Gamma_{\omega N}(\rho_{B0}),~c_{\omega}$ can be obtained
\begin{gather}
       \Gamma_{\omega N}(\rho_{B0})=\frac{m_\omega}{\rho_{B0}}\sqrt{2(E/A+M_N)\rho_{B0}-m_\sigma^2\sigma_0^2-2\left(\varepsilon_{\rm kin}^n+\varepsilon_{\rm kin}^p\right)},\\
       \omega_0=\frac{\Gamma_{\omega N}(\rho_{B0})\rho_{B0}}{m_\omega^2},\\
       c_\omega=\frac{b_\omega(d_\omega+1)-y\left[1+b_\omega(1+d_\omega)^2\right]}{(d_\omega+1)+y\left[1+b_\omega(1+d_\omega)^2\right](d_\omega+1)^2},
\end{gather}
where
$$
y=\frac{\frac{1}{2}m_\sigma^2\sigma_0^2-\frac{1}{2}m_\omega^2\omega_0^2-P_{\rm kin}^p-P_{\rm kin}^n}{2\rho_{B0}\Gamma_{\omega N}(\rho_{B0})\omega_0}.
$$
$\Gamma_{\rho N}(\rho_{B0})$ and $a_{\rho}$ can be calculated numerically with the definition of $E_{\rm sym}$ and $L$ with Eqs. \eqref{eqn.esym} and \eqref{eqn.L} simultaneously and constants $b_\omega,~d_\omega$ can be obtained by solving Eqs. \eqref{eqn.K} and \eqref{eqn.J} simultaneously. Finally, we can have $a_\omega$ with Eq. \eqref{eqn.aw}. The obtained parameters mentioned above, $g_{\sigma}^{q},~\Gamma_{\omega N}(\rho_{B0}),~a_{\omega},~b_{\omega},~c_{\omega},~d_{\omega},~\Gamma_{\rho N}(\rho_{B0}),~a_{\rho}$, are listed in Table \ref{table.para}.

\begin{table}[H] 
    \centering 
    \caption{The coupling parameters, $g_{\sigma}^q,~\Gamma_{\omega N}(\rho_{B0}),~a_{\omega},~b_{\omega},~c_{\omega},~d_{\omega}$, $\Gamma_{\rho N}(\rho_{B0}),~a_{\rho}$, obtained by fitting saturation properties in Table \ref{table.SNMsat}. }\label{table.para}
    \begin{tabular}{c|c|cccccccccccccccc}
    \hline \hline
       $M_{N0}^*/M_N$ ~~& $J_0$[MeV] ~~&$g_{\sigma}^{q}$ ~~& $\Gamma_{\omega N}(\rho_{B0})$ ~~& $a_{\omega}$ ~~& $b_{\omega}$ ~~&$c_{\omega}$ ~~&$d_{\omega}$  ~~& $\Gamma_{\rho N}(\rho_{B0})$ ~~& $a_{\rho}$ \\
       \hline    
       \multirow{2}*{0.556} 
       & -800 &\multirow{2}*{6.4516} &\multirow{2}*{15.0304} 
       & 1.2838 & 0.1306 & 0.2801 & 0.5887
       &\multirow{2}*{7.2479} & \multirow{2}*{0.4755} \\ 
       & 400 & & & 1.1955 & 0.2398 & 0.4284 & 0.1743  & & \\
       \cline{1-10}
       \multirow{2}*{0.70} 
       & -800 &\multirow{2}*{4.4885}  & \multirow{2}*{10.9401} 
       & 1.0269 & 0.04339 & 0.007140 & 0.0001920 
       &\multirow{2}*{8.12239} & \multirow{2}*{0.4150} \\
       & 400 & & & 1.0156 & 0.5852 & 0.6504 & -0.4724 & & \\
       \hline\hline
    \end{tabular}
\end{table} 

\begin{figure}[htbp]
	\centering
	\includegraphics[scale=0.5]{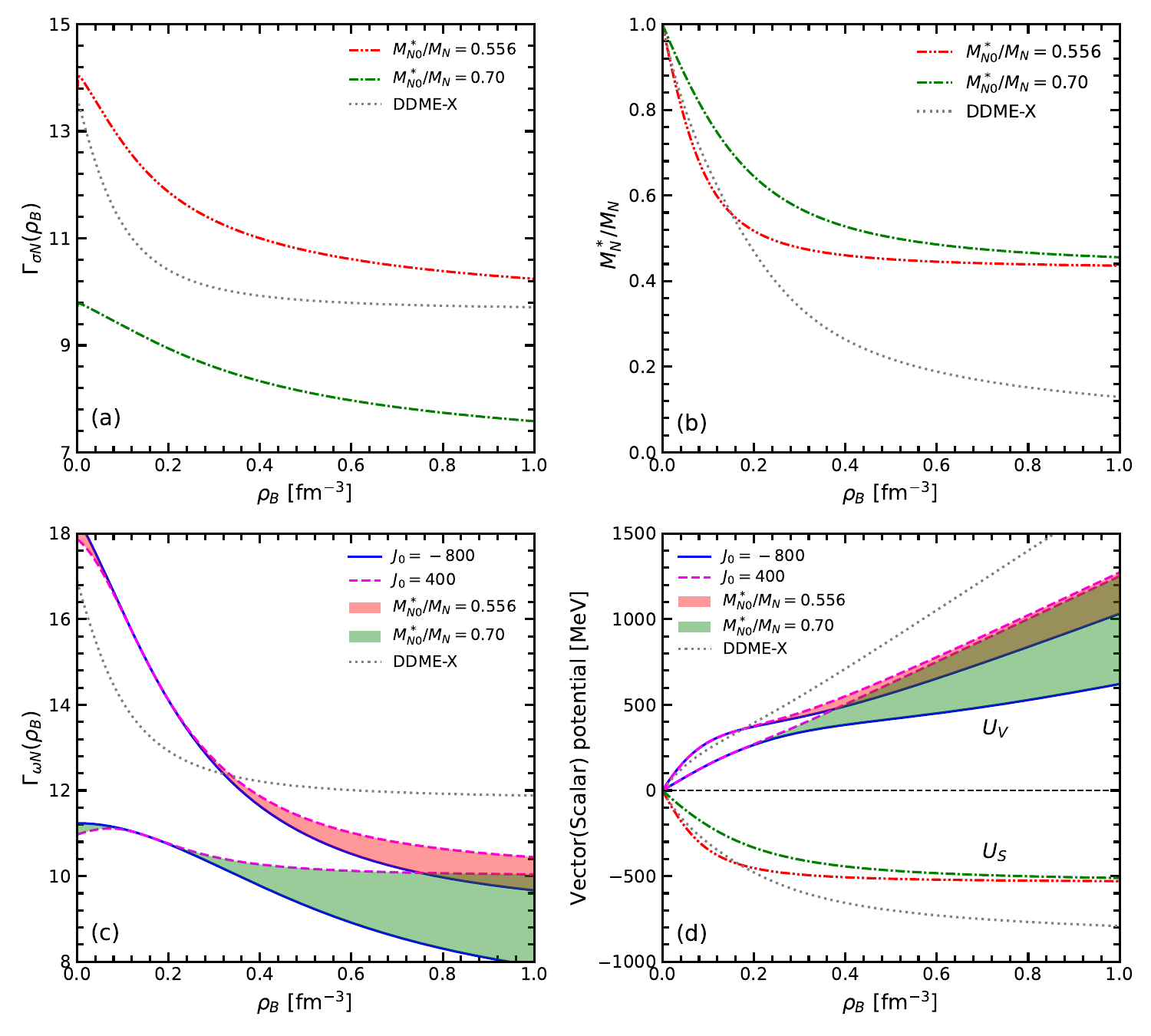}
	\caption{The density-dependent behaviors of coupling constants, $\Gamma_{\sigma N}(\rho_B)$ and $\Gamma_{\omega N}(\rho_B)$, the effective mass, $M_N^*$, and the scalar(vector) potential at $M_{N0}^*/M_N=0.556,~0.70$ as functions of the baryon density for symmetric nuclear matter with DDQMF parameters in Table \ref{table.para}. }\label{fig.coup}
\end{figure}   

The density-dependent behaviors of coupling constants, $\Gamma_{\sigma N}(\rho_B)$ and $\Gamma_{\omega N}(\rho_B)$, the effective mass, $M_N^*$, and the scalar(vector) potential for symmetric nuclear matter with $M_{N0}^*/M_N=0.556,~0.70$ are plotted in the panels of Fig. \ref{fig.coup}, respectively.  In the panel (a), the effective $\sigma$ meson couplings constants, $\Gamma_{\sigma N}(\rho_B)=M^*_N/\sigma$ are given as a function of baryon density. The strengths of those with $M_{N0}^*/M_N=0.556$ are larger than the ones with $M_{N0}^*/M_N=0.70$ around $40\%$. The $\Gamma_{\sigma N}(\rho_B)$ from DDME-X are compared, which are smaller than the lower effective mass case in DDQMF model, although their nucleon effective mass at saturation density is the same.  In panel (b), the effective nucleon masses are given as functions of baryon density.  The DDQMF parameterization with $M_{N0}^*/M_N=0.556$ has similar density-dependent behaviors with DDME-X below the nuclear saturation density. When the density increases, the effective mass from DDME-X rapidly reduces while the ones in DDQMF tend to converge at high densities.

Correspondingly, the vector meson coupling constants are shown in panel (c) which depend on the skewness $J$ in the fitting process. The red and green shaded regions in this panel with the upper limit marked by the magenta dashed line corresponds to $J_0 = 400$ MeV and the blue solid line to $J_0 = -800$ MeV. The brownish regions in this paper result from the overlap of green and red regions. The relevant results from the DDME-X model are also added for comparison. $\Gamma_{\omega N}(\rho_B)$ with $M_{N0}^*/M_N=0.556$ will be stronger than that with $M_{N0}^*/M_N=0.70$ to provide more repulsive contributions, which will be canceled out with the attraction from the $\sigma$ meson. At the low-density region, the $\Gamma_{\omega N}(\rho_B)$ with $M_{N0}^*/M_N=0.556$ is larger than that from DD-MEX, while it will be smaller above $\rho_{B}=0.3$ fm$^{-3}$. Therefore, EOSs from the DDQMF will be softer than that from DD-MEX at the high-density region due to the lack of strong repulsion. Furthermore,  $\Gamma_{\omega N}(\rho_B)$ with larger $M_{N0}^*$ are more sensitive to $J_0$. When $J_0=-800$ MeV, it rapidly decreases with the baryon density.  The scalar and vector potentials, $U_S$ and $U_V$, from DDME-X and DDQMF models are plotted in panel (d), which are strongly related to effective nuclear mass and vector meson coupling strength. Therefore, they have similar behaviors as shown in panels (b) and (c).
 
\begin{figure}[htbp]
	\centering
	\includegraphics[scale=0.5]{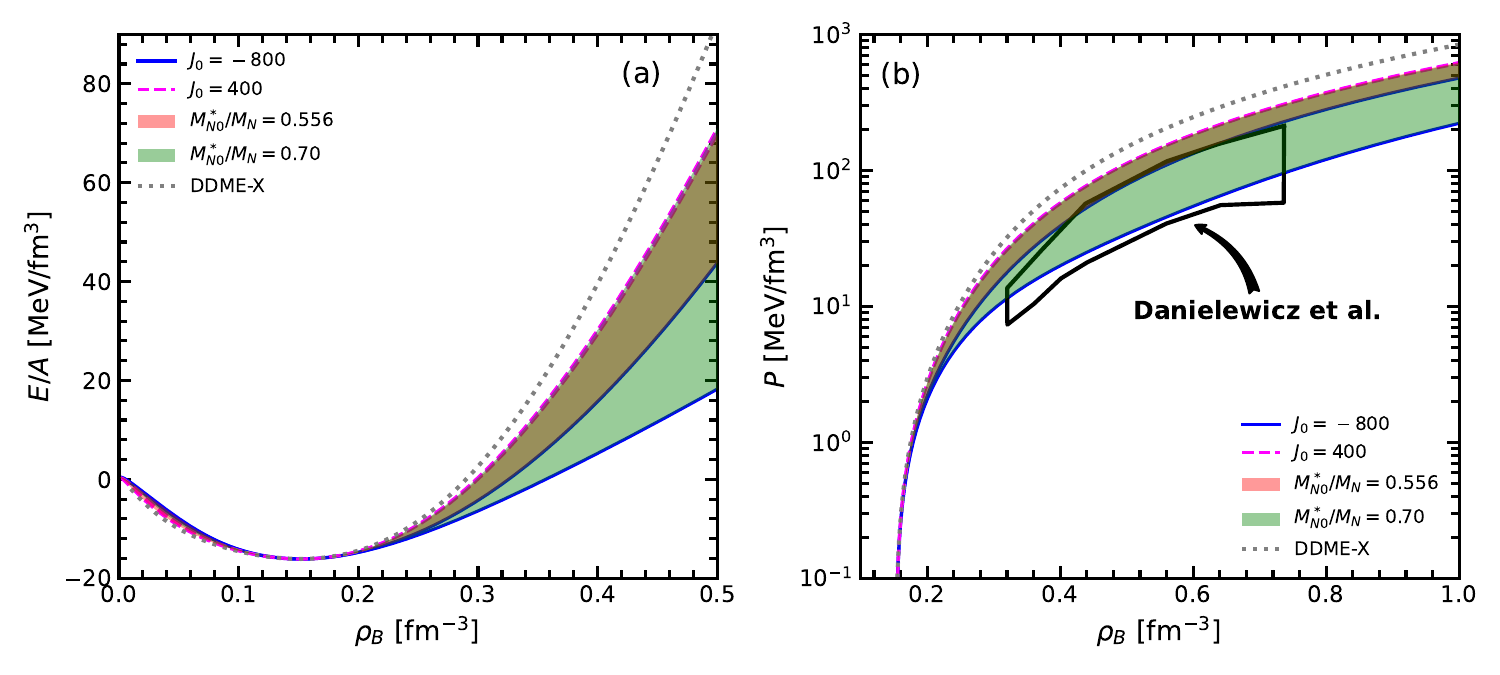}
	\caption{(a) The binding energies per nucleon and (b) pressures as functions of vector density for symmetric nuclear matter with DDQMF parameters in Table \ref{table.para}.}\label{fig.SNMeos}
\end{figure}

In Fig. \ref{fig.SNMeos}, we show the behaviors of the binding energies per nucleon, $E/A$, and pressures, $P$ as functions of baryon density for symmetric nuclear matter in panels (a) and (b). When the $M_{N0}^*/M_N$ is fixed, the EOS will become softer with smaller $J_0$, since this skewness term denotes the third-order derivative of $E/A$ from Eq. \eqref{eqn.J}. Meanwhile, for the smaller $J_0$, the DDQMF with a smaller effective nucleon mass can generate a stiffer EOS due to the larger vector meson contributions. However, when $J_0$ is large enough, such as $J_0 = 400$ MeV, the effect of $M_{N0}^*$ becomes very weak and two EOSs from different effective masses are almost the same. So the magenta dashed lines, which represent the case of $J_0 = 400$ MeV, for $M_{N0}^*/M_N=0.556,~0.70$ seem to be almost overlapping. Furthermore, the EOS from DDME-X is stiffer than those from DDQMF due to its strong repulsive vector potential. In panel (b), the pressures from the present model are compared to the constraint from heavy-ion collisions at $2-4\rho_{B0}$ \cite{danielewicz2002}, which supports the DDQMF parameterization with a larger $M_{N0}^*$ and smaller $J_0$. Similarly, the pressures for $J_0 = 400$ MeV at $M_{N0}^*/M_N=0.556,~0.70$ are almost the same.

 Together with $\beta$-equilibrium and charge neutrality conditions in Eq. \eqref{eqn.betaequip}, the EOSs of neutron star matter with the DDQMF model can be obtained. The EOS of the nonuniform matter in the crust region is generated by IUFSU parameterization with Thomas-Fermi approximation from Ref. \cite{bao2014}, where the crust EOSs with different $L_{\rm crust}=47$ MeV and $L_{\rm crust}=110$ MeV are given for comparison. They denote the neutron skin measurements from CREX and PREX-II, respectively. 
 
\begin{figure}[htbp]
 	\centering
 	\includegraphics[scale=0.5]{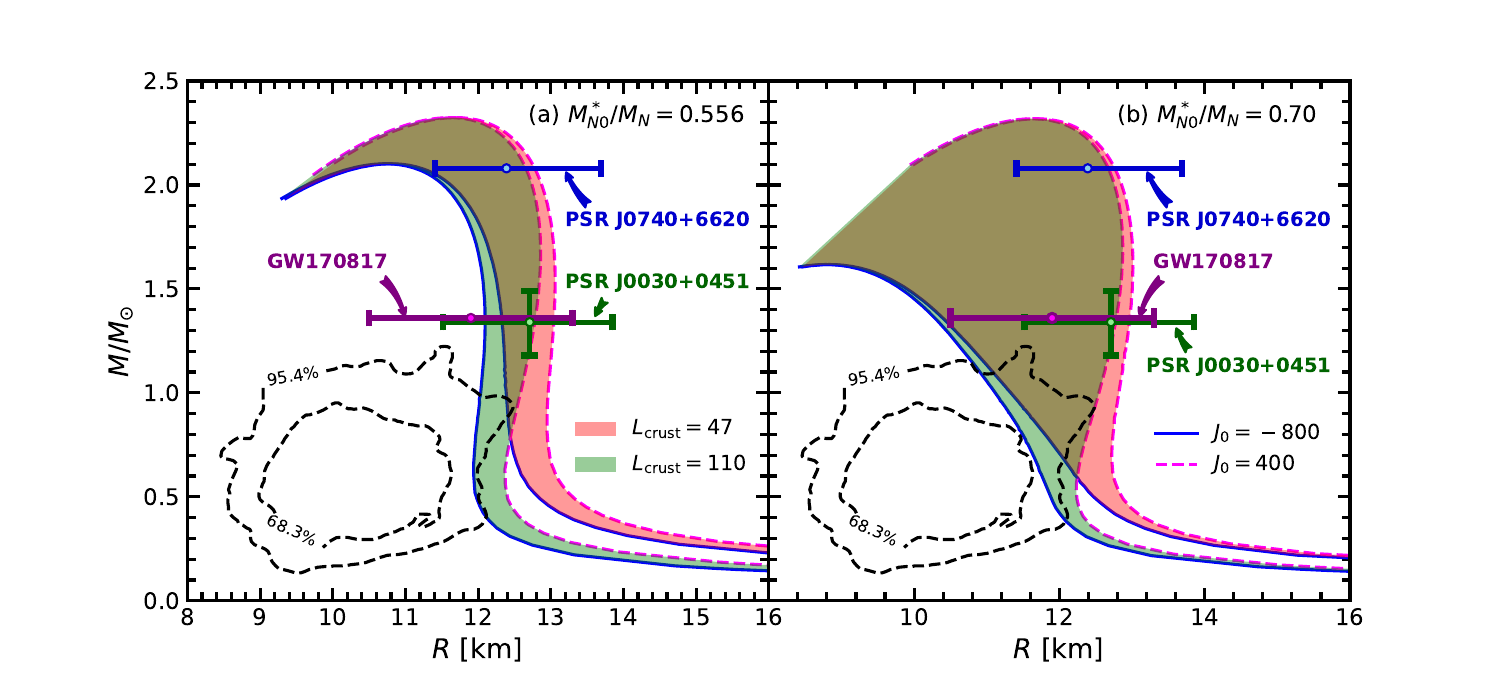}
 	\caption{Mass-radius relations of neutron stars at $M_{N0}^*/M_N=0.556,~0.70$ with the DDQMF model. }\label{fig.MR}
 \end{figure}

 The mass-radius ($M-R$) relation of neutron stars can be calculated by solving the Tolman–Oppenheimer–Volkoff (TOV) equation \cite{tolman1939,oppenheimer1939} with the EOSs of neutron star matter as input. In panels (a) and (b) of Fig. \ref{fig.MR}, the $M-R$ relations at $M_{N0}^*/M_N=0.556,~0.70$ with different skewness, $J_0$ are shown, respectively. Additionally, we include mass-radius observations from measurements of PSR J0030+0451 \cite{miller2019} and PSR J0740+6620 \cite{miller2021} by NICER, which have a mass of $1.34^{+0.15}_{-0.16}M_\odot$ with a radius $12.71^{+1.24}_{-1.06}$ km and a mass of $2.072^{+0.067}_{-0.066}M_\odot$ with a radius $12.39^{+1.30}_{-0.98}$ km, respectively. The purple horizontal line indicates the radius constraint at 1.4$M_\odot$ from GW170817 event with $R_{\rm 1.4}=11.9\pm 1.4$ km \cite{abbott2017}. The mass-radius constraints from the compact central object of HESS J1731-347 \cite{doroshenko2022} are also shown with $68\%$ and $95\%$ confidence intervals. It should be noted that the red and green shaded regions in this figure are calculated from $-800\leq J_0\leq 400$ MeV with distinguished crust EOSs with $L_{\rm crust}=47$ MeV and $L_{\rm crust}=110$ MeV, respectively, which are different from the meanings represented in the previous figure. The dashed line and solid line still represent the upper and lower limit, and the brownish region results from the overlap of green and red region. Apparently, a higher $L_{\rm crust}$ can yield a softer EOS, leading to a smaller radius in the low-mass region. 
 
 The maximum masses for $M_{N0}^*/M_N=0.556$ and $M_{N0}^*/M_N=0.70$ are all about $M_{\rm max}=2.3M_\odot$ with radius about $R_{\rm max}=11.7$ km at $J_0=400$ MeV, while they are much different at $J_0=-800$ MeV. For the DDQMF with $M_{N0}^*/M_N=0.556$ and $J_0=-800$ MeV, the maximum mass of the neutron star is about $2.1M_\odot$. It largely decreases to $1.6M_\odot$ for $M_{N0}^*/M_N=0.70$.  The $M-R$ relation for $M_{N0}^*/M_N=0.556$ with $L_{\rm crust}=110$ MeV can satisfy the $95\%$ credibility constraint from HESS J1731-347 as well as the constraints from PSR J0740+6620, PSR J0030+0451 and GW170817 event. On the other hand, for $M_{N0}^*/M_N=0.70$, the $M-R$ relations are more sensitive to $J_0$, where the maximum masses and the corresponding radius, as well as the radius at the low-mass region, at $J_0=-800$ MeV are much smaller. The $M-R$ relation for $M_{N0}^*/M_N=0.70$ at $J_0=-800$ MeV with $L_{\rm crust}=110$ MeV can even approach the 68\% credibility constraint from HESS J1731-347, but can not satisfy the constraints from the other three observations about the massive neutron star. 
 
\begin{figure}[htbp]
 	\centering
 	\includegraphics[scale=0.5]{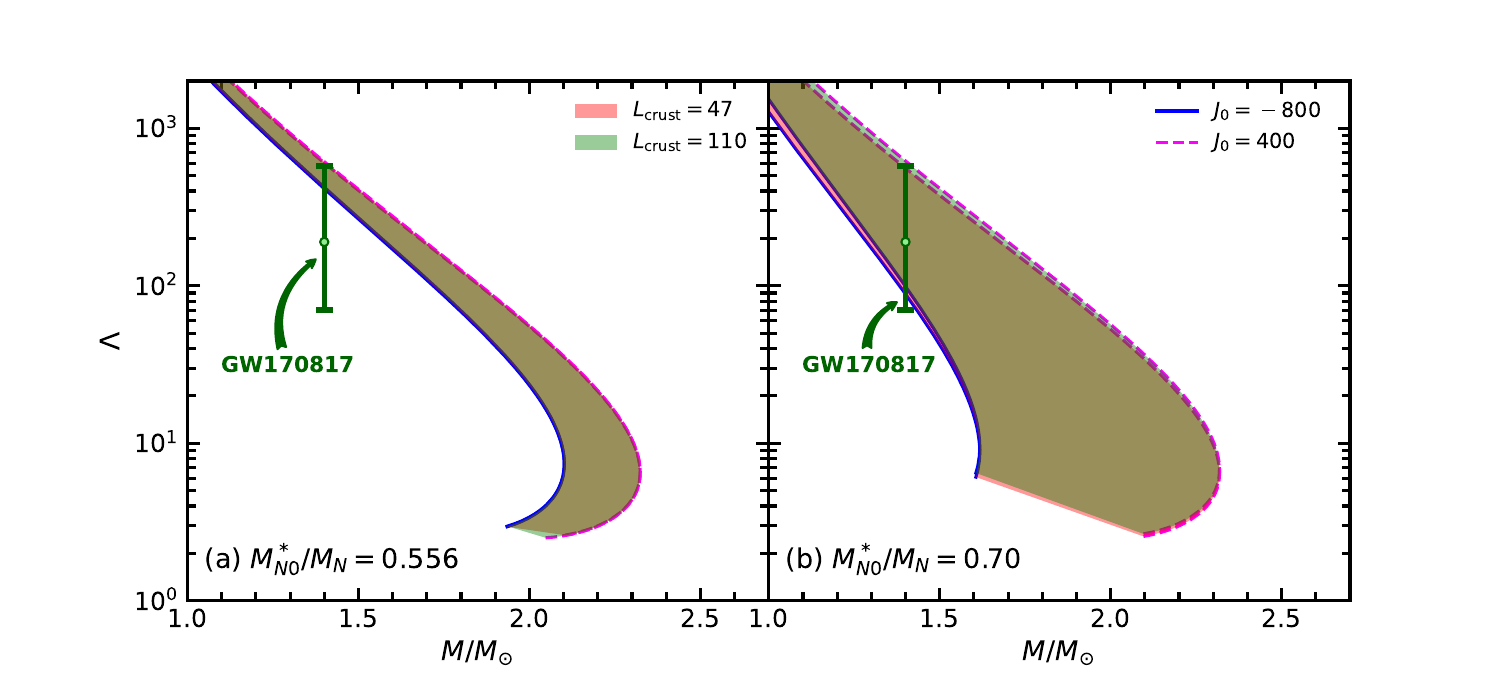}
 	\caption{Tidal deformabilities of neutron stars at $M_{N0}^*/M_N=0.556,~0.70$ with the DDQMF model.}\label{fig.MLam}
\end{figure}

 With the rapid development of gravitational wave detectors, the dimensionless tidal deformability, $\Lambda$, has also become another important property of neutron stars to constrain the theoretical models. The GW170817 event provides the constraint on $\Lambda$ at $M_{1.4}$ with $\Lambda_{1.4}=190^{+390}_{-120}$ \cite{abbott2018}. In Fig. \ref{fig.MLam}, $\Lambda$ of the neutron stars as a function of their masses from DDQMF models are shown.  For $M_{N0}^*/M_N=0.556$, $\Lambda_{1.4}$ with $L_{\rm crust}=47-110$ MeV change from $410-420$ at $J_0=-800$ MeV to $602-612$ at $J_0=400$ MeV. Similarly, for $M_{N0}^*/M_N=0.70$, $\Lambda_{1.4}$ change from $90-100$ at $J_0=-800$ MeV to $558-620$ at $J_0=400$ MeV. Therefore, $\Lambda_{1.4}$ from these DDQMF models can almost satisfy the constraint from GW170817 event. Similar to the behavior of the mass, for $M_{N0}^*/M_N=0.70$, the $\Lambda$ is more sensitive to $J_0$, and $\Lambda$ is much smaller for the same mass at $J_0=-800$ MeV, so the constraint from GW170817 event supports a larger $M_{N0}^*$ and a smaller $J_0$, which is consistent with the requirements of heavy-ion collisions.
 
By adjusting the $\rho$ meson coupling constants, $\Gamma_{\rho N}(\rho_{B0})$ and $a_\rho$, using Eq. \eqref{eqn.esym} and Eq. \eqref{eqn.L}, we can obtain the core EOSs of the DDQMF model with different $L$ values, while maintaining the symmetry energy fixed at sub-saturation density $\rho_{B}=0.11~ \rm fm^{-3}$ with $E_{\rm sym}(0.11)=26.85$ MeV for $M_{N0}^*/M_N=0.556$ and $E_{\rm sym}(0.11)=26.99$ MeV for $M_{N0}^*/M_N=0.70$. The magnitudes of $\Gamma_{\rho N}$ and $a_\rho$ for $L=30,~40,~60,~80$ MeV are listed in Table \ref{table.para_diffL}. We excluded the case below $L = 30$ MeV because the corresponding speed of sound in nuclear matter becomes less than zero. 

\begin{table}[htbp]
	\centering 
	\caption{Parameters $\Gamma_{\rho N}$ and $a_\rho$ of the DDQMF model generated for different $L_0$ at saturation density $\rho_{B0}$ with the symmetry energy $E_{\rm sym}$ fixed at $\rho_{B0}=0.11~\rm fm^{-3}$.}\label{table.para_diffL}
	\begin{tabular}{c|c|cccccccc}
	\hline \hline    
	&$L_0$ [MeV]    & 30~~      & 40~~       & 60~~      & 80~~     \\
	\hline    
	\multirow{2}*{$M_N^*/M_N=0.556$} 
	&$\Gamma_{\rho N}(\rho_{B0})$    & 6.69347~~ & 6.99212~~  & 7.49302~~ & 7.91292  \\ 
	&$a_\rho$                        & 0.76350~~ & 0.60553~~  & 0.35513~~ & 0.15780  \\ 
	\hline    
	\multirow{2}*{$M_N^*/M_N=0.70$}   
	&$\Gamma_{\rho N}(\rho_{B0})$    & 7.65896~~ & 7.90445~~  & 8.33656~~ & 8.71322  \\ 
	&$a_\rho$                        & 0.63759~~ & 0.51342~~  & 0.32079~~ & 0.16086  \\ 
	\hline \hline 
	\end{tabular}
\end{table}

\begin{figure}[htbp]
	\centering
	\includegraphics[scale=0.5]{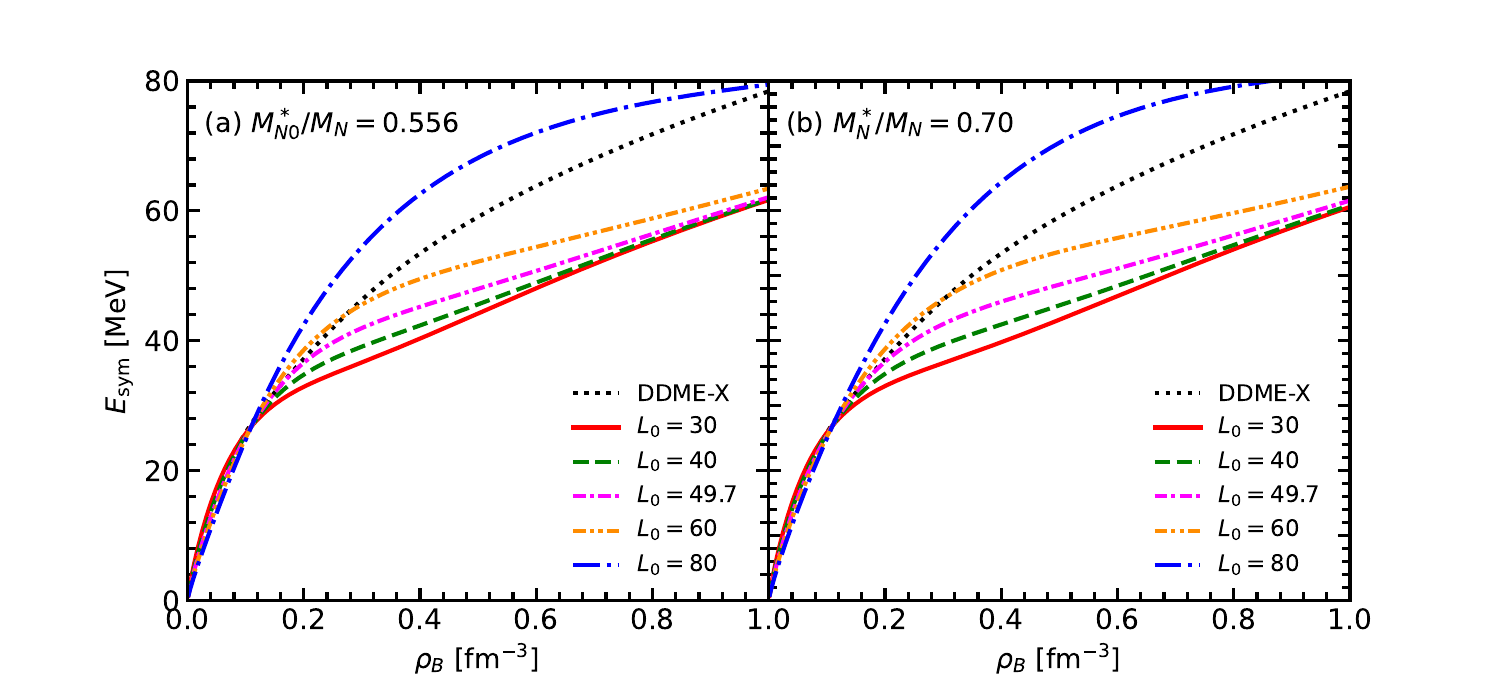}
	\caption{The density-dependent behaviors of the symmetry energy for $L_0=30,~40,~60,~80$ MeV of the DDQMF model.}\label{fig.coup_differL}
\end{figure}

The density-dependent behaviors of $E_{\rm sym}$ are plotted in Fig. \ref{fig.coup_differL}. Smaller $L$ can yield larger $E_{\rm sym}$ below the sub-saturation density, and produce smaller $E_{\rm sym}$ in the high-density region, which can be understood by the expansion of $E_{\rm sym}(\rho_B) = E_{\rm sym}(\rho_0) + L(\rho_B - \rho_0)/3\rho_0 + \cdots $. The $E_{\rm sym}$ will converge at the high density due to the second term related to  $\Gamma_{\rho N}(\rho_{B})$ will disappear at the high density. $E_{\rm sym}$ for $L_0=30-60$ MeV will converge at about 1.0 $\rm fm^{-3}$ while $E_{\rm sym}$ for $L_0=80$ MeV will converge at about 2.0 $\rm fm^{-3}$. The effective mass will sightly affect the symmetry energy for $L_0=80$ MeV and a larger effective mass generates a larger symmetry energy at high density.

\begin{figure}[htbp]
	\centering
	\includegraphics[scale=0.5]{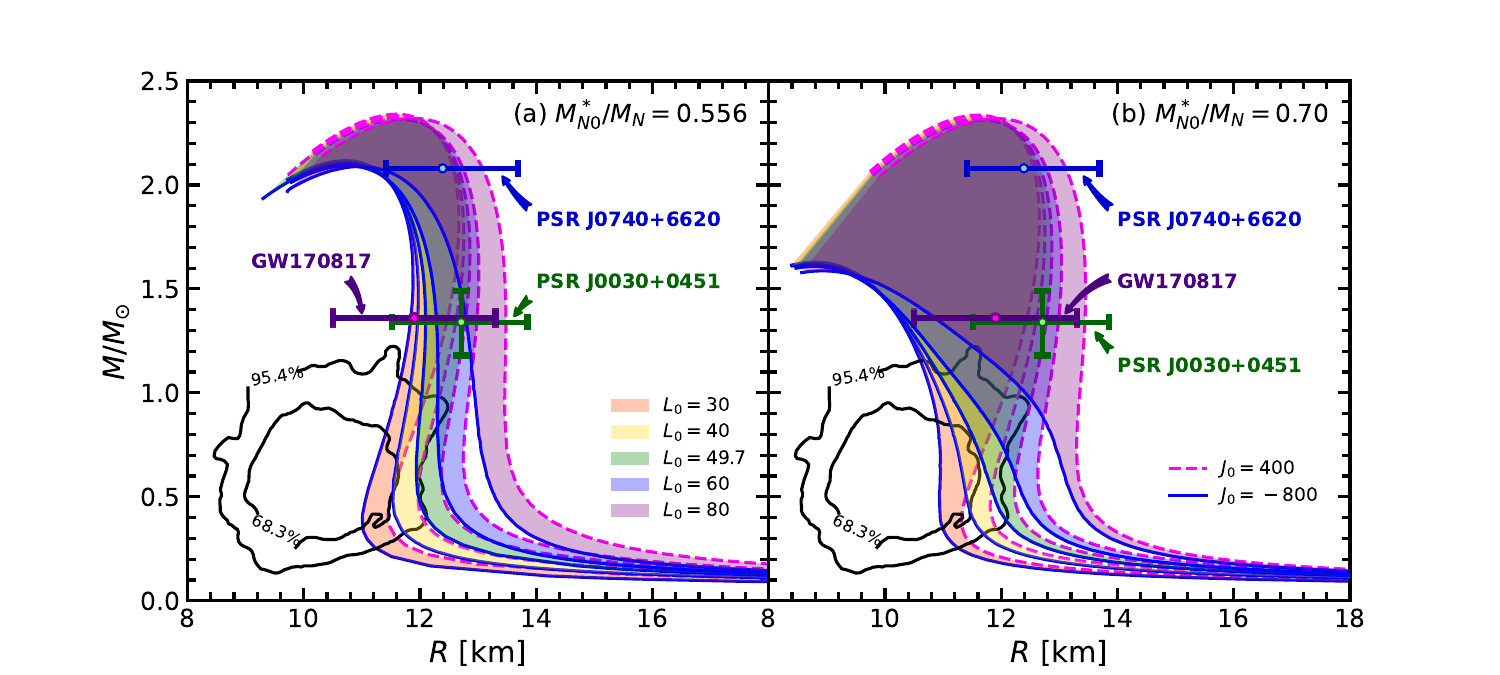}
	\caption{Mass-radius relations of neutron stars with different $L_0$ for the core EOS with the DDQMF model.}\label{fig.MR_differL}
\end{figure}

Now we combine the core EOS with different $L_0$ in DDQMF model with $L_{\rm crust}=110$ MeV for the crust EOS from IUFSU model since a larger $L_{\rm crust}$ can yield a smaller radius, especially in the low-mass region, which can make it easier to satisfy the $M-R$ constraint from HESS J1731-347. The corresponding $M-R$ relations with the above EOSs in Table \ref{table.para_diffL} are shown in Fig. \ref{fig.MR_differL}. The colorful shadow regions relate to the core EOSs with different $L_0$, and $M-R$ relation with the original $L_0=49.7$ MeV is also shown for comparison. We can find that the $L_0$ for the core EOS has little effect on the maximum mass and the corresponding radius of neutron stars.

The $M-R$ relations from different $L_0,~J_0$ with $M_N^*/M_N=0.556$ are given in panel (a) of Fig. \ref{fig.MR_differL}. The radius at $1.4 M_\odot$, $R_{1.4}$, decreases from $(12.76,~13.48)$ km to $(11.86,~12.50)$ km, a decrease of about $1.0$ km as the core EOS changes from $L_0=80$ MeV to $L_0=30$ MeV, while the radius at  $0.77 M_\odot$ ($R_{0.77}$) decreases by about $1.7$ km, from $(13.03,~13.49)$ km to $(11.38,~11.81)$ km, where $(R_{J_0=-800},~R_{J_0=400})$ denotes the radius interval from $J_0=-800$ MeV to $J_0=400$ MeV at a fixed effective nucleon mass. This indicates that the $L_0$ of the core EOS plays an opposite role in determining the radius of low-mass neutron stars compared to $L_{\rm crust}$ of the crust EOS. Furthermore, the $M-R$ relation obtained from $L_0=30$ MeV (shaded in red) can fully satisfy the $68\%$ credibility $M-R$ constraint from HESS J1731-347, as well as the mass-radius constraints from PSR J0030+0451, PSR J0740+6620 and GW170817 event. We also show the similar $M-R$ relations for $M_N^*/M_N=0.70$ in panel (b). As $J_0$ approaches $400$ MeV, the $M-R$ relation for $L_0=30$ MeV may also satisfy all the $68.3\%$ credit constraints and the radius constraint from GW170817 event. When the skewness becomes smaller, the massive neutron star cannot be supported.
  
\begin{figure}[htbp]
	\centering
	\includegraphics[scale=0.5]{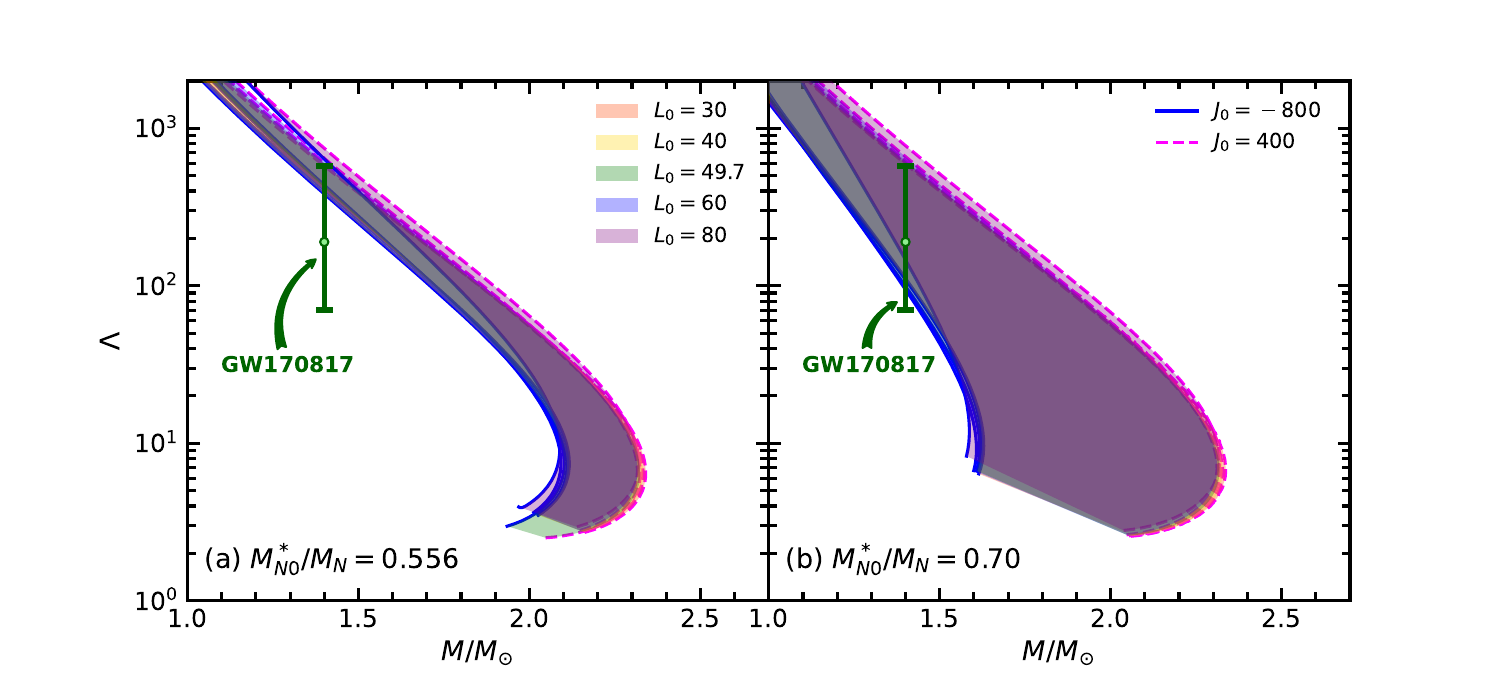}
	\caption{Tidal deformabilities of neutron stars with different $L_0$ for the core EOS with the DDQMF model.}\label{fig.MLam_differL}
\end{figure}

Finally, $\Lambda$ of the neutron stars as a function of their masses from the EOSs of different $L_0,~J_0$ with $M_N^*/M_N=0.556,~0.70$ given in Fig. \ref{fig.MR_differL} are shown in Fig. \ref{fig.MLam_differL}.  For $M_{N0}^*/M_N=0.556$, $\Lambda_{1.4}$ with $L_0=30-80$ MeV change from $405-640$ at $J_0=-800$ MeV to $575-745$ at $J_0=400$ MeV. Similarly, for $M_{N0}^*/M_N=0.70$, $\Lambda_{1.4}$ with $L_0=30-80$ MeV change from $110-145$ at $J_0=-800$ MeV to $610-780$ at $J_0=400$ MeV. $\Lambda_{1.4}$ from these DDQMF models with different $L_0$ can almost satisfy the constraint from GW170817 event except for the case of $L_0=80$ MeV at $M_{N0}^*/M_N=0.556$, which is the purple region in the panel (a) of Fig. \ref{fig.MLam_differL}, since larger $L_0$ for the core EOS can produce stiffer EOS of the neutron star matter and the larger $\Lambda$ will be obtained.

\section{Conclusion}\label{sec4}
In this work, we developed the density-dependent quark mean-field (DDQMF) model, where the $\omega,~\rho$ meson coupling constants and the nonlinear terms in the previous QMF model \cite{xing2016} were replaced with density-dependent coupling constants. Seven independent parameters  $(g_{\sigma}^q,~\Gamma_{\omega N}(\rho_{B0}),~b_{\omega},~c_{\omega},~d_{\omega},~\Gamma_{\rho N}(\rho_{B0}),~a_{\rho})$ were determined by fitting the nuclear saturation properties ($\rho_{B0},~E/A,~K_0,~J_0,~E_{\rm sym 0},~L_0,~M_{N0}^*$) of nuclear matter from DDME-X model, where the effective nucleon mass was fixed by two cases with $M_{N0}^*/M_N=0.556,~0.70$ because present core-collapse supernova simulation favors a larger $M_{N0}^*$ \cite{yasin2020} while $M_{N0}^*/M_N\sim 0.60$ can give reasonable spin-orbit splittings for finite nuclei in RMF model. $J_0$ was chosen to be in the range of $-800\leq J_0\leq 400$ MeV based on the analysis of terrestrial nuclear experiments and energy density functional theory \cite{zhang2018}. 

We investigated the properties of infinite nuclear matter and neutron stars with the DDQMF model. The larger $M_N^*$ corresponds to the smaller vector potential, which will provide a softer EOS and make it easier to satisfy the constraint from heavy-ion collisions at $2-4\rho_{B0}$ densities. However, when $J_0$ is large enough, e.g. $J_0=400$ MeV, the EOSs with $M_{N0}^*/M_N=0.556$ and $M_{N0}^*/M_N=0.70$ are almost the same, which leads to the maximum mass of neutron stars of around $M_{\rm max}=2.32M_\odot$ with radius about $R_{\rm max}=11.7$ km for both $M_{N0}^*$ at $J=400$ MeV. Moreover, the $M-R$ relations obtained from $M_{N0}^*/M_N=0.556$ can simultaneously satisfy the constraints from PSR J0740+6620, PSR J0030+0451 and GW170817 event. However, the $M-R$ relations obtained from $M_{N0}^*/M_N=0.70$ at very small $J_0$ can hardly satisfy these three constraints since the EOSs and $M-R$ relations with larger $M_N^*$ will be more sensitive to $J_0$, and the maximum mass and the corresponding radius, as well as the radius at the low-mass region, from $M_{N0}^*/M_N=0.70$ at $J_0=-800$ MeV are much lower than those at $J_0=400$ MeV. Furthermore, when the crust EOS is chosen to be $L_{\rm crust}=110$ MeV, $M-R$ relations from $M_{N0}^*/M_N=0.70$ at $J_0=-800$ MeV can even come close to satisfying the $68\%$ credibility constraint from HESS J1731-347.

To further study the effect of $L_0$ on the properties of neutron stars, we obtained several core EOSs with different $L_0$ by adjusting $\Gamma_{\rho N}(\rho_{B0})$ and $a_{\rho}$ to control the strength of $\rho N$ interaction, while keeping $E_{\rm sym}$ fixed at $\rho_{B}=0.11~ \rm fm^{-3}$. We found that $L_0$ has minimal impact on the maximum mass and the corresponding radius, while the radius for low-mass neutron star becomes smaller with $L_0$ decreasing, which is opposite to the effect of the $L_{\rm crust}$. Furthermore, for $M_{N0}^*/M_N=0.556$, the $M-R$ relation obtained by combining the core EOS with $L_0=30$ MeV and the softer crust EOS with $L_{\rm crust}=110$ MeV can fully satisfy the $68\%$ credibility $M-R$ constraint from HESS J1731-347, as well as the mass-radius constraints from PSR J0030+0451, PSR J0740+6620 and GW170817. However, for $M_N^*/M_N=0.70$, the $M-R$ relation for $L_0=30$ MeV can satisfy all these constraints only if $J_0$ is close to $400$ MeV. In addition, the tidal deformabilities at $M_{1.4}$, $\Lambda_{1.4}$, from these DDQMF models can almost satisfy the constraint, $\Lambda_{1.4}=190^{+390}_{-120}$, from GW170817 event except for the case of $L_0=80$ MeV at $M_{N0}^*/M_N=0.556$. Therefore, the constraint from GW170817 event supports a larger $M_{N0}^*$ and a smaller $J_0$.

In the DDQMF model, the number of parameters in density-dependent coupling constants is reduced compared to the DDRMF model since the effective nucleon mass is generated from the quark level. Furthermore, the density-dependent behaviors of the coupling constants in DDQMF model at the high-density region also have obvious differences from the DDRMF model, which can provide a constraint from the nucleon internal structure. We will apply the DDQMF model to study the finite nuclei system and introduce the high-momentum correlation to discuss the effect of nucleon structure on the properties of nuclei in the future. 
 
In the inner core region of a neutron star, baryons including strangeness degrees of freedom, such as $\Lambda,~\Sigma$ and $\Xi$ hyperons, will be present when the Fermi energies of nucleons are larger than hyperon free masses which is also called as a hyperonic star. In our previous work, we studied the properties of hyperonic stars using the DDRMF model \cite{huang2022}, and additionally, many works have taken hyperons into account in the framework of QMF model \cite{shen2002,hu2014prc,xing2017,hu2017}. So, we will apply the DDQMF model to study the properties of hyperonic matter and the hyperonic star in future. Furthermore, we will perform a Bayesian analysis on the parameters of DDQMF model with the proper prior ranges obtained in the present work.

\section{Acknowledgments}
This work was supported in part by the National Natural Science Foundation of China	 (Grant No. 12175109).

\end{document}